\documentclass[%
superscriptaddress,
preprint,
showpacs,preprintnumbers,
 showkeys,
 amsmath,amssymb,
 aps,
plb,
]{revtex4-1}

\usepackage{latexsym}
\usepackage{psfrag}
\usepackage{graphics}
\usepackage{graphicx}
\usepackage{makeidx}
\usepackage{color}
\usepackage{amsfonts}
\usepackage{dcolumn} % Align table columns on decimal point
\usepackage{bm} % bold math
\usepackage{setspace,supertabular}
\usepackage{lineno}
 \usepackage{subfigure}
\usepackage{appendix}
\usepackage{graphicx}% Include figure files
\usepackage{dcolumn}% Align table columns on decimal point
\usepackage{bm}% bold math
\begin{document}

\newcommand{\ket}[1]{{\vert #1 \rangle}}
\newcommand{\bra}[1]{{\langle #1 \vert}}
\newcommand{\brak}[2]{{\langle #1\vert #2 \rangle}}
\def\sls #1{\rlap/\kern  -  .1em #1}
\newcommand{\di}{\displaystyle}
\renewcommand{\vec}[1]{\mathbf{#1}}

\title{Relativistic Elastic Differential Cross Sections for Equal Mass Nuclei}% Force line breaks with \\

\author{C. M. Werneth}
\affiliation{%
 NASA Langley Research Center, 2 West Reid Street, Hampton, VA 23681
}
\author{K. M. Maung}%
\affiliation{%
The University of Southern Mississippi, 118 College Drive, Box 5046, Hattiesburg, MS 39406 
}
\author{W. P. Ford}%
\affiliation{%
The University of Southern Mississippi, 118 College Drive, Box 5046, Hattiesburg, MS 39406 
}

\date{\today}% It is always \today, today,

\begin{abstract}
%The minimum requirement for studying relativistic effects in the Lippmann-Schwinger equation is the inclusion of relativistic kinematics. It is demonstrated %that the relativistic (REL) and non-relativistic (NR) propagator share similar pole-structure for projectiles and targets of equal mass for heavy-ion %scattering. Utilizing the first order optical potential, the REL and NR on-shell scattering amplitudes are shown to be indistinguishable near the elastic cut. %As a result, no differences are observed between NR and REL elastic differential cross sections for equal mass systems. 
The effects of relativistic kinematics are studied for nuclear collisions of equal mass nuclei.
It is found that the relativistic and non-relativistic elastic scattering amplitudes are nearly indistinguishable, and, hence, the relativistic and non-relativistic differential cross sections become indistinguishable.  
These results are explained by analyzing the Lippmann-Schwinger equation with the first order optical potential that was employed in the calculation.

%\begin{description}
%\end{description}
\end{abstract}


\maketitle

The Lippmann-Schwinger (LS) equation for the transition amplitude is often used to calculate nuclear cross sections \cite{Joachain}. 
The minimum requirement for the study of relativistic effects in the LS equation is the inclusion of relativistic kinematics.
In this letter, the effect of relativistic (REL) kinematics in nucleus-nucleus scattering is studied, and it is found that the scattering amplitudes calculated with REL and non-relativistic (NR) kinematics are nearly indistinguishable for nuclear collisions of equal mass nuclei when using the first order optical potential.  It is shown that there are no observed significant differences between the NR and REL elastic differential cross sections for the equal mass case.

The elastic scattering amplitude is determined from the transition amplitude, which is obtained by solving the following integral equation \cite{Joachain}, 
\begin{equation}
T_{\rm AA}(\vec k', \vec k) = U(\vec k', \vec k) + \int \frac{U(\vec k', \vec k'') T_{\rm AA}(\vec k'', \vec k)}{E_k - E_{k''} + i \epsilon} d \vec k'',
\end{equation}
where $\vec k$ ($\vec k'$) is the initial (final) relative momentum of the projectile-target system in the center of momentum (CM) frame, $E$ is the NR or REL total energy, $T_{\rm AA}(\vec k', \vec k)$ is the off-shell transition amplitude, and $ U(\vec k', \vec k)$ is the optical potential.  Using factorization and on-shell approximations for central potentials, the optical potential for projectiles and targets of equal mass is expressed as \cite{Pickle,Pickle2, Wolfe, NASATP2014, WernethElArch}
\begin{equation}
U(\vec k', \vec k) =  \eta {\rm A_P A_T}  t_{\rm NN}(|\vec k' - \vec k|)\rho_{\rm A_P} (|\vec k'- \vec k|) \rho_{\rm A_T}(|\vec k' - \vec k|),
\end{equation}
where $\eta$ is the M\"{o}ller factor \cite{Moller, Joachain}, $\rm A_P$ and $\rm A_T$ are, respectively, the number of nucleons in the projectile and target, 
$t_{\rm NN}$ is the nucleon-nucleon (NN) transition amplitude, and $\rho(|\vec k' - \vec k|)$ is the nuclear density of a nucleus \cite{DeVries1,DeVries2}. For equal mass projectile and target nuclei ($\rm A_P = A_T=A)$, $\eta = 1$, and the optical potential is
\begin{equation}
U(\vec k', \vec k) =  {\rm A}^2 \rho^2_{\rm A}(|\vec k'- \vec k|)t_{\rm NN}(|\vec k' - \vec k|).
\end{equation}

The on-shell scattering amplitude is related to the on-shell transition amplitude by \cite{Joachain}
\begin{equation}
f(k,k, \hat {\vec k} \cdot \hat {\vec k'}) = -(2 \pi)^2 k \frac{dk}{dE_k} t(k, k, \hat {\vec k'} \cdot \hat {\vec k}). \label{onshell}
\end{equation}
The off-shell scattering amplitude is defined as,
\begin{equation}
f(\vec k', \vec k) \equiv \beta(k')~ t(\vec k', \vec k) ~\beta(k), \label{ftot}
\end{equation}
where
\begin{equation}
\beta (k) = 2 \pi i \sqrt{k\frac{dk}{dE_k} } \label{beta},
\end{equation}
such that equation \eqref{onshell} is satisfied when the relative momentum is on-shell. 

In order to show that the NR and REL elastic differential cross sections are the same for equal mass systems, the off-shell scattering amplitudes are obtained by using equation \eqref{ftot};
\begin{align}
\beta_{\rm AA}(k') T_{\rm AA} (\vec k', \vec k) \beta_{\rm AA}(k) &=  \frac{\beta_{\rm AA}(k')}{\beta_{\rm NN}(\kappa')} \beta_{\rm NN}(\kappa') t_{\rm NN}(|\vec k' - \vec k|)  \beta_{\rm NN}(\kappa) \frac{ \beta_{\rm AA}(k) } {\beta_{\rm NN} (\kappa) } A^2 \rho_A^2(|\vec k' - \vec k|) \label{eq:amp}\\
&+  {\bf P}\int \Bigg[  \beta_{\rm AA}(k') \frac{\beta_{\rm NN} (\kappa ')} { \beta_{\rm NN} (\kappa ')} t(|\vec k' - \vec k''|)  
\frac{\beta_{\rm NN} (\kappa '')} { \beta_{\rm NN} (\kappa '')} A^2 \rho^2_A(|\vec k' - \vec k''|)  \nonumber \\
 & ~~~~~~~~~\times \frac{1}{E_k - E_{k''}} \frac{ \beta_{\rm AA}(k'') }{  \beta_{\rm AA}(k'')}  T_{\rm AA} (| \vec k''- \vec k|) \beta_{\rm AA}(k) d \vec k'' \Bigg] \nonumber \\
&-i \pi \int \Bigg[ \beta_{\rm AA}  (k') \frac{ \beta_{\rm NN}(\kappa ')} { \beta_{\rm NN}(\kappa ') } t(|\vec k' - \vec k|) 
 \frac{ \beta_{\rm NN}(\kappa '')} { \beta_{\rm NN}(\kappa '') }
A^2 \rho_A^2(|\vec k' - \vec k''|) \nonumber \\
&~~~~~~~~\times \delta \Big( E_k - E_{k''} \Big) \frac{ \beta_{\rm AA}(k'') } {  \beta_{\rm AA}(k'')     }
T_{\rm AA} (\vec k'', \vec k) \beta_{\rm AA}(k) d \vec k'' \Bigg] \nonumber, 
\end{align}
where
\begin{equation}
\beta^2_{\rm AA} =
  \begin{cases}
     -(2 \pi )^2 \frac{{\rm A}m}{2}  & \text{for the NR case}, \label{BetaAA} \\
		-(2 \pi)^2 {\rm A} \frac{ \sqrt{m^2 + \kappa^2}} {2} & \text{for the REL case}
  \end{cases}
\end{equation}

\begin{equation}
\beta^2_{\rm NN} =
  \begin{cases}
     -(2 \pi )^2 \frac{m}{2}  & \text{for the NR case} \\
		-(2 \pi)^2  \frac{ \sqrt{m^2 + \kappa^2}} {2} & \text{for the REL case}, \label{BetaNN}
  \end{cases}
\end{equation}
and the propagator has been expressed in terms of its principal value, ${\bf P}$, 
\begin{equation}
\frac{1}{E_k - E_{k''} + i \eta} = {\bf P} \Bigg( \frac{1}{E_k - E_{k''}}\Bigg) - i \pi \delta(E_k - E_{k''}).
\end{equation}
The Fermi motion of the nucleons inside the nucleus is neglected; therefore, the momentum imparted to each nucleon is $\kappa = k/{\rm A}$. 
The mass of the nucleus, $M$, is approximately ${\rm A} m$, where $m$ is the average nucleon mass. 

The pole structure for NR scattering amplitude is proportional to $1/(k^2-{k''}^2)$.  With REL kinematics, the propagator can be rationalized such that the pole structure is manifestly the same as the NR case, thus
\begin{equation}
{\bf P} \Bigg( \frac{1}{E_k - E_{k''}}\Bigg) =  \label{Pvalue}
\begin{cases}
{\bf P} \Bigg[ \frac{2\mu}{(k^2 - {k''}^2)} \Bigg] & \text{for the NR case}
\\
{\bf P} \Bigg[ \Bigg( \frac{\sqrt{M^2 + k^2}}{2} \Bigg) \Bigg( \frac{ 1 + \sqrt{\frac{M^2 + {k''}^2}{M^2 + {k}^2} }  }{k^2 - {k''}^2 } \Bigg) \Bigg]   & \text{for the REL case},
\end{cases}
\end{equation}
where $\mu = {\rm M}/2$.

Using equations \eqref{BetaAA}, \eqref{BetaNN}, and \eqref{Pvalue}, the scattering amplitude from equation \eqref{eq:amp} reduces to the following for on-shell scattering:
\begin{align}
F_{\rm AA} &(k, \hat {\vec k'} \cdot \hat {\vec k}) = f_{\rm NN} (k, \hat {\vec k'} \cdot \hat {\vec k}){\rm A}^3 \rho_A(k, \hat {\vec k'} \cdot \hat {\vec k}) \label{eqREL} \\
&+ {\bf P} \int \frac{ f_{\rm NN} (|\vec k' - \vec k'' |) h(k,k'')
{\rm A}^3 \rho_{\rm A}^2(|\vec k' - \vec k''|)  F_{\rm AA} (| \vec k'' - \vec k|) }{[-(2\pi)^2]  (k^2 - {k''}^2)} d \vec k'' \nonumber \\
&-i \pi \int f_{\rm NN}(k, \hat {\vec k'} \cdot \hat {\vec k''})  {\rm A}^3 \rho_{\rm A}^2(k, \hat {\vec k'} \cdot \hat {\vec k''}) F_{\rm AA} (k, \hat {\vec {k''} } \cdot \hat {\vec k}) \frac{k}{-(2 \pi)^2} d \Omega_{k''} \nonumber
\end{align}
where
\begin{equation}
\begin{cases}
h(k,k'') = 2 & \text{for the NR case} \\
 h(k,k'') = 1 + \sqrt{   \frac{M^2 + {k''}^2 }{  M^2 + k^2  }   } & \text{for the REL case}.
\end{cases}
\end{equation}
The only difference between the two amplitudes is that $h(k,k'') = 2$ in the NR case, and $h(k,k'') \rightarrow 2$ only near the elastic cut for the REL case. 
The optical potential is largest near the on-shell momentum but decays rapidly thereafter.  By definition, the principal value integral is never evaluated at $k'' = k$; however, significant contributions occur \emph{near} the elastic cut. Due to the rapidly decaying optical potential and large contributions near the elastic cut, little differences are observed between the NR and REL scattering amplitudes for projectiles and targets of equal mass.

To illustrate these results, elastic differential cross sections are given in Fig. 1 for $^4$He + $^{56}$Fe, $^4$He + $^{20}$Ne,  $^4$He + $^{12}$C, and $^4$He + $^{4}$He reactions at a lab projectile energy of 1 GeV/n. NR and REL elastic differential cross sections are generated with a three-dimensional Lippmann-Schwinger (LS3D) solution \cite{LS3D_1,LS3D_2,LS3D_3,LS3D_4}. See reference \cite{NASATP2014} for the explicit form of the nuclear densities, transition amplitude, and parameterizations that were used in the current work. 

\begin{figure*}
\begin{center}
\subfigure{\includegraphics[scale=0.33]{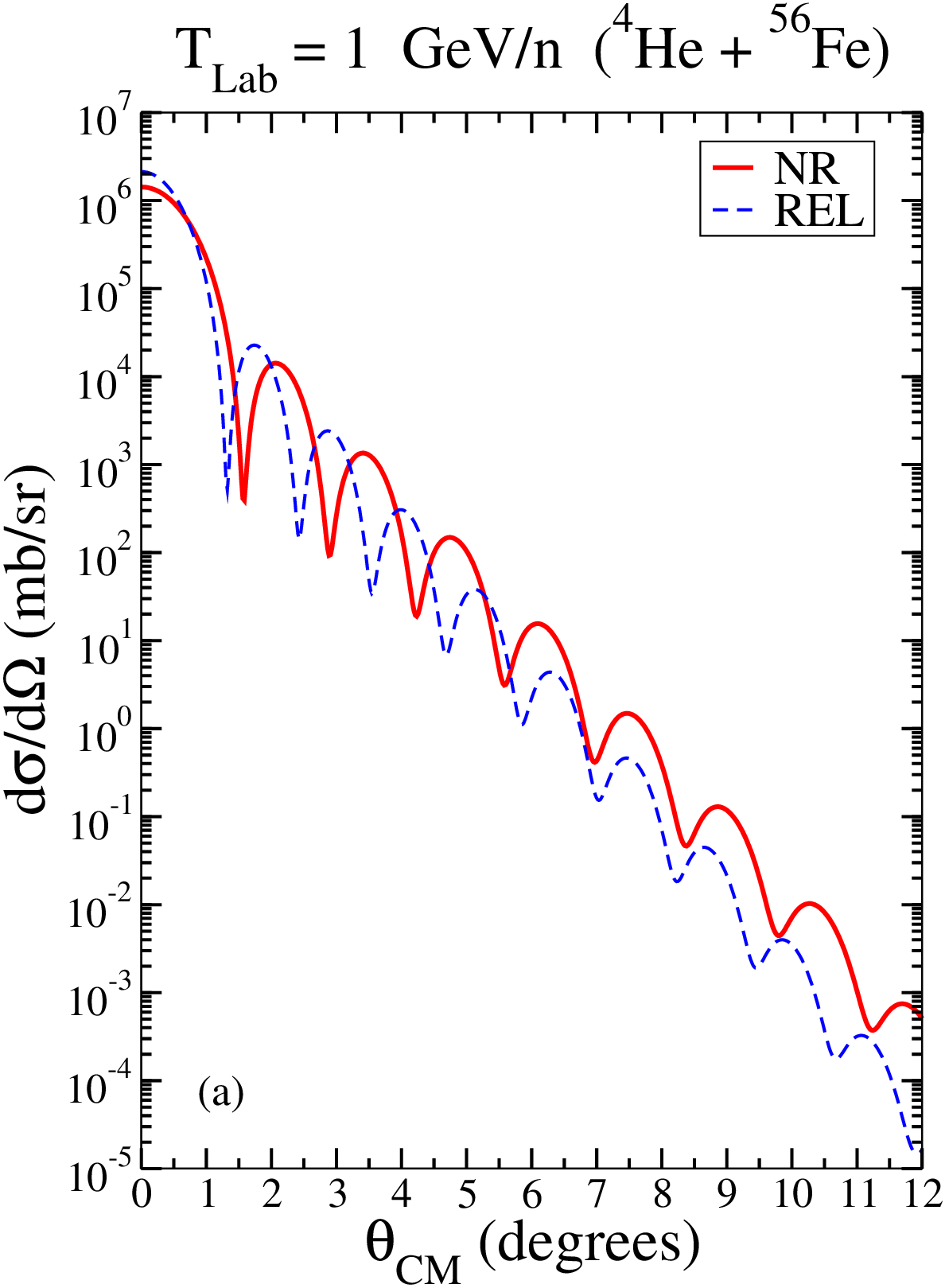}} \hspace{0.5cm}
\subfigure{\includegraphics[scale=0.33]{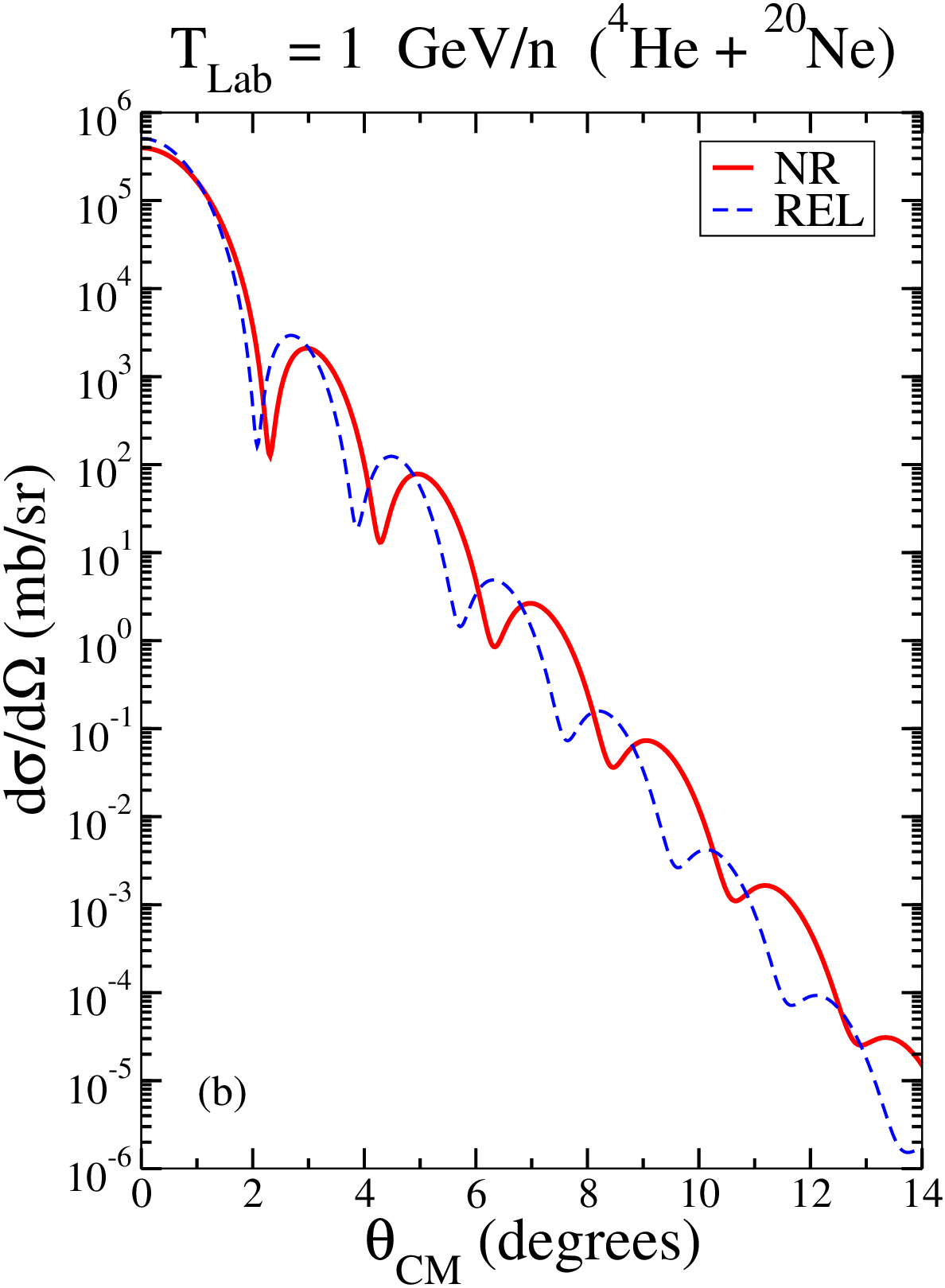}}  \\
\subfigure{\includegraphics[scale=0.33]{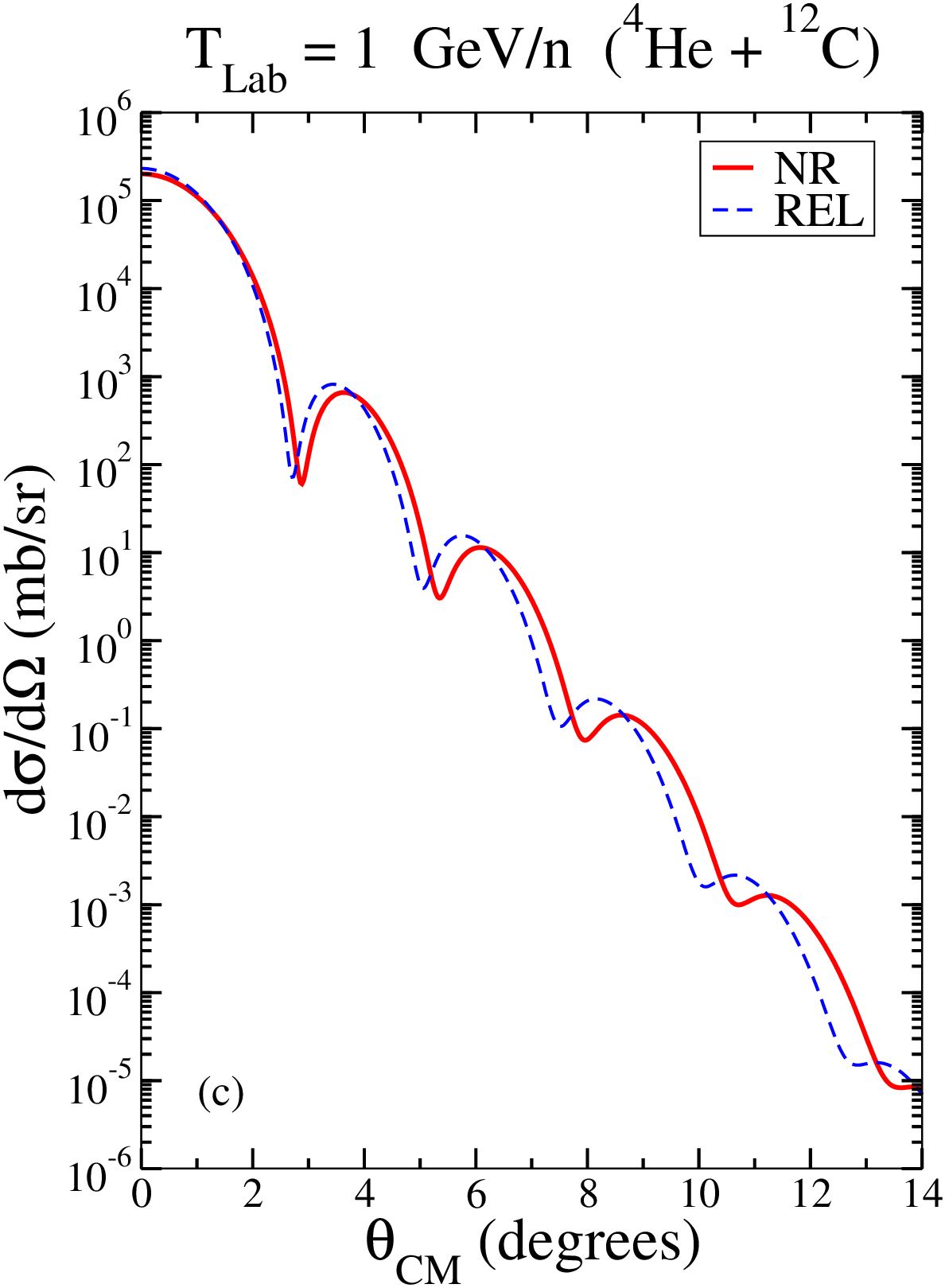}} \hspace{0.5cm}
\subfigure{\includegraphics[scale=0.33]{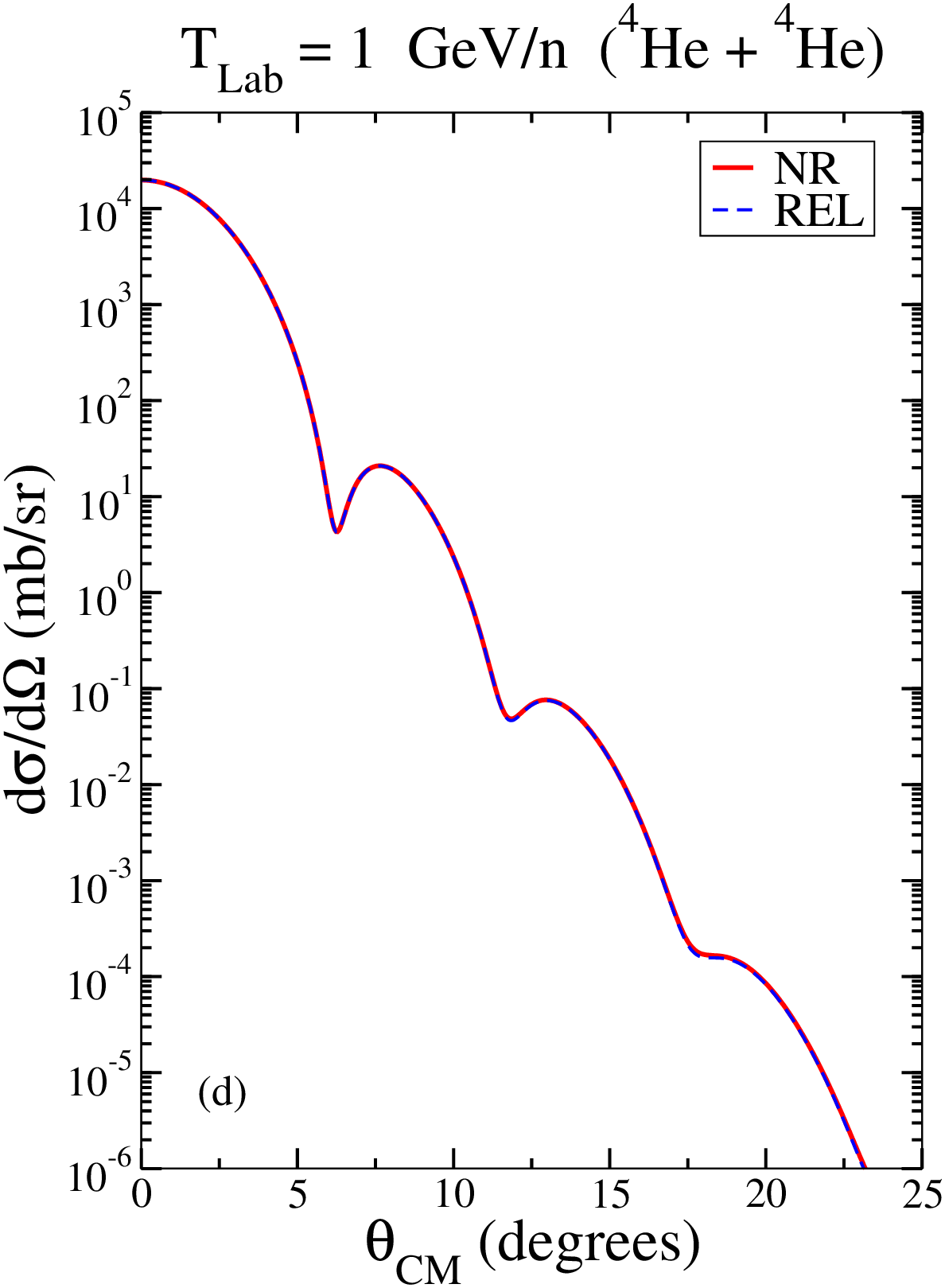}} 
\caption[Elastic differential cross sections...]{Elastic differential cross sections for $^4$He + $^{56}$Fe, $^4$He + $^{20}$Ne,  $^4$He + $^{12}$C, and $^4$He + $^{4}$He reactions at a lab projectile energy of 1 GeV/n. NR results are indicated with solid red lines, and REL results are given as dashed blue lines.}
\label{HeReactions}
\end{center}
\end{figure*}

From Fig. \ref{HeReactions}, it is obvious that the NR and REL elastic differential cross sections are different for the $^4$He + $^{56}$Fe, $^4$He + $^{20}$Ne, and $^4$He + $^{12}$C reactions, where the projectile and target masses differ. It is also observed that the largest REL difference occurs in the case of $^4$He + $^{56}$Fe, where the mass difference between the projectile and target is largest. There are no significant differences between the NR and REL results for the equal mass case of the $^4$He + $^{4}$He reaction.

\begin{figure*}
\begin{center}
\subfigure{\includegraphics[scale=0.33]{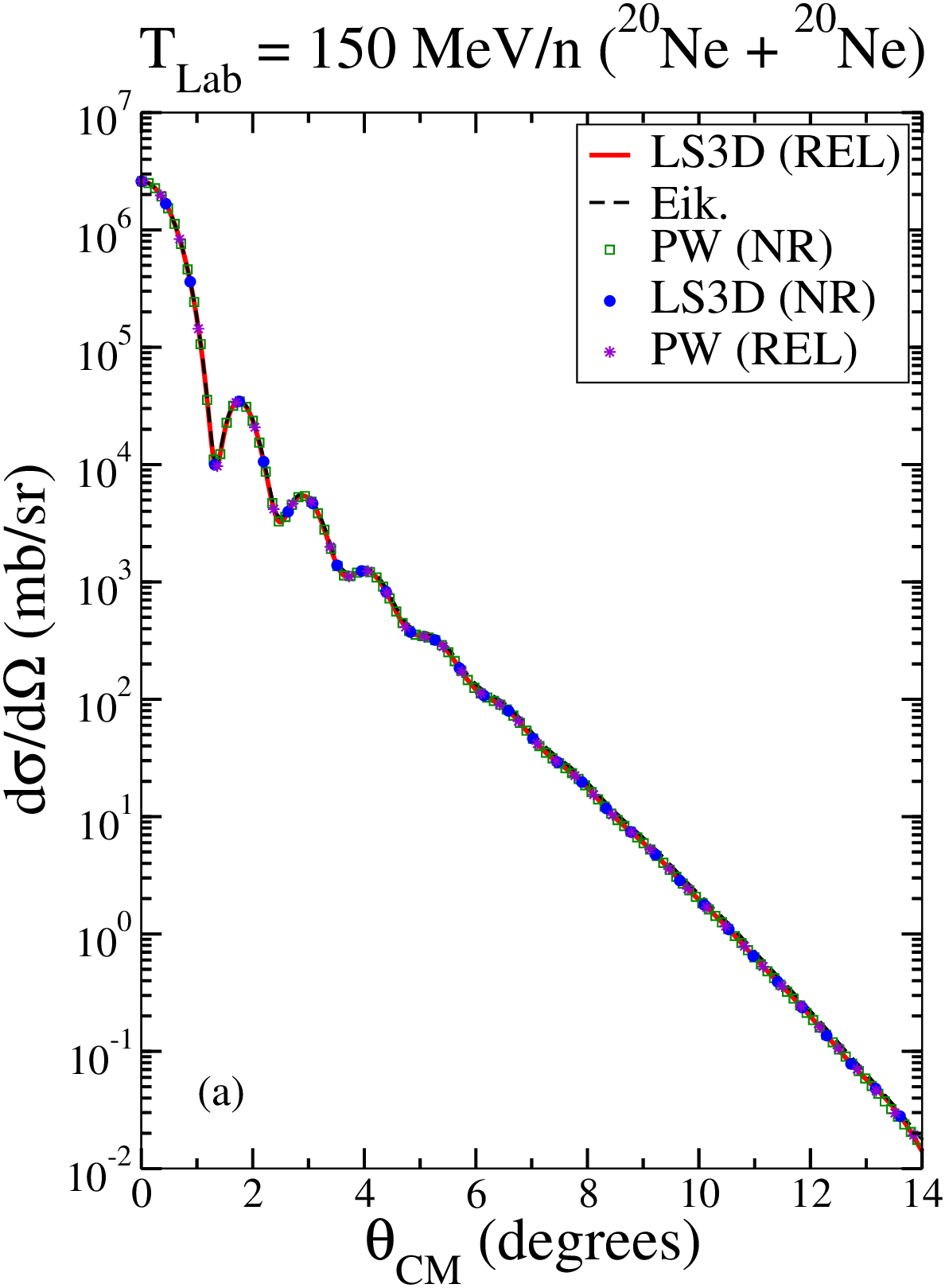}} \hspace{0.5cm}
\subfigure{\includegraphics[scale=0.33]{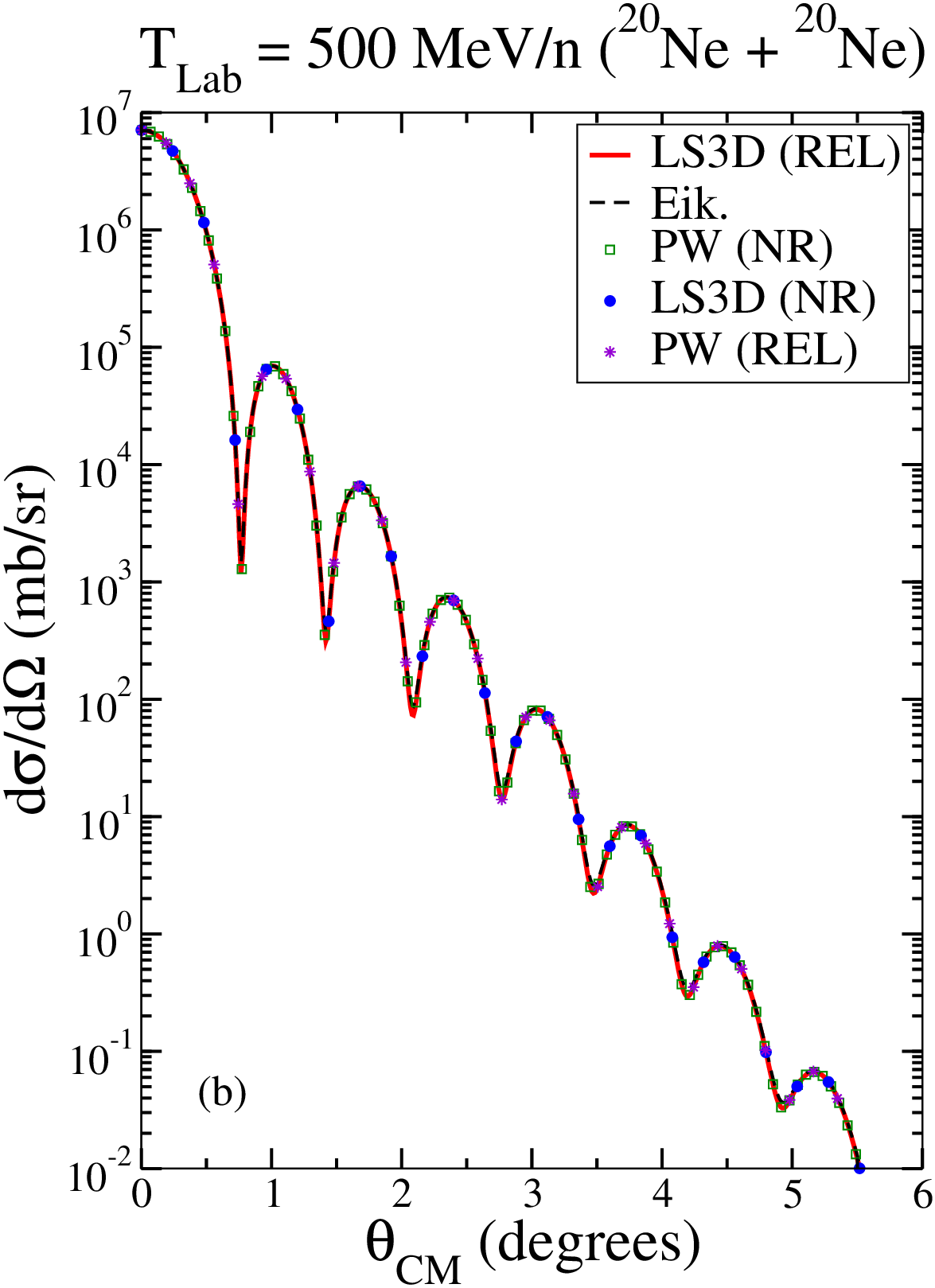}}  \\
\subfigure{\includegraphics[scale=0.33]{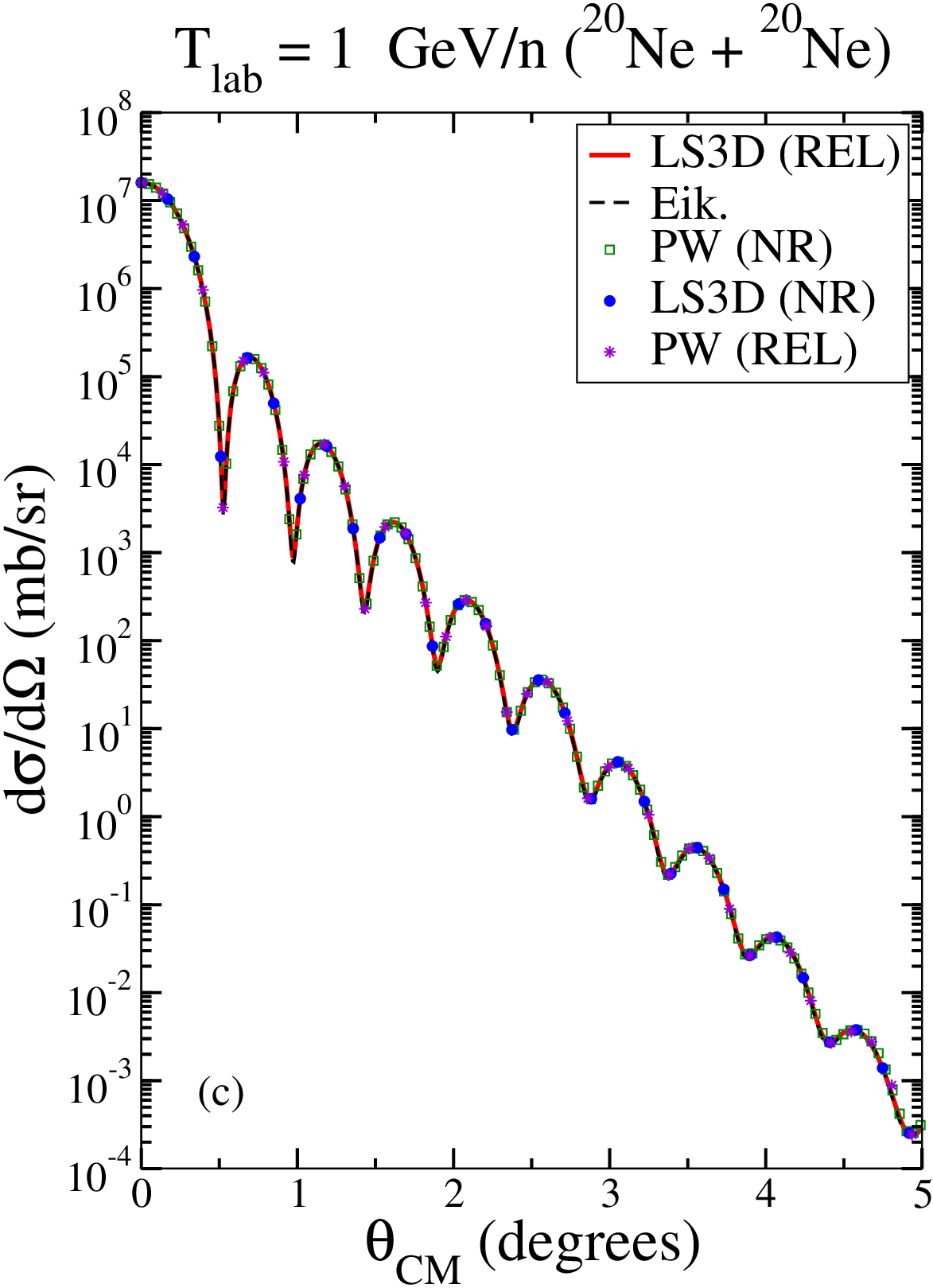}} \hspace{0.5cm}
\subfigure{\includegraphics[scale=0.33]{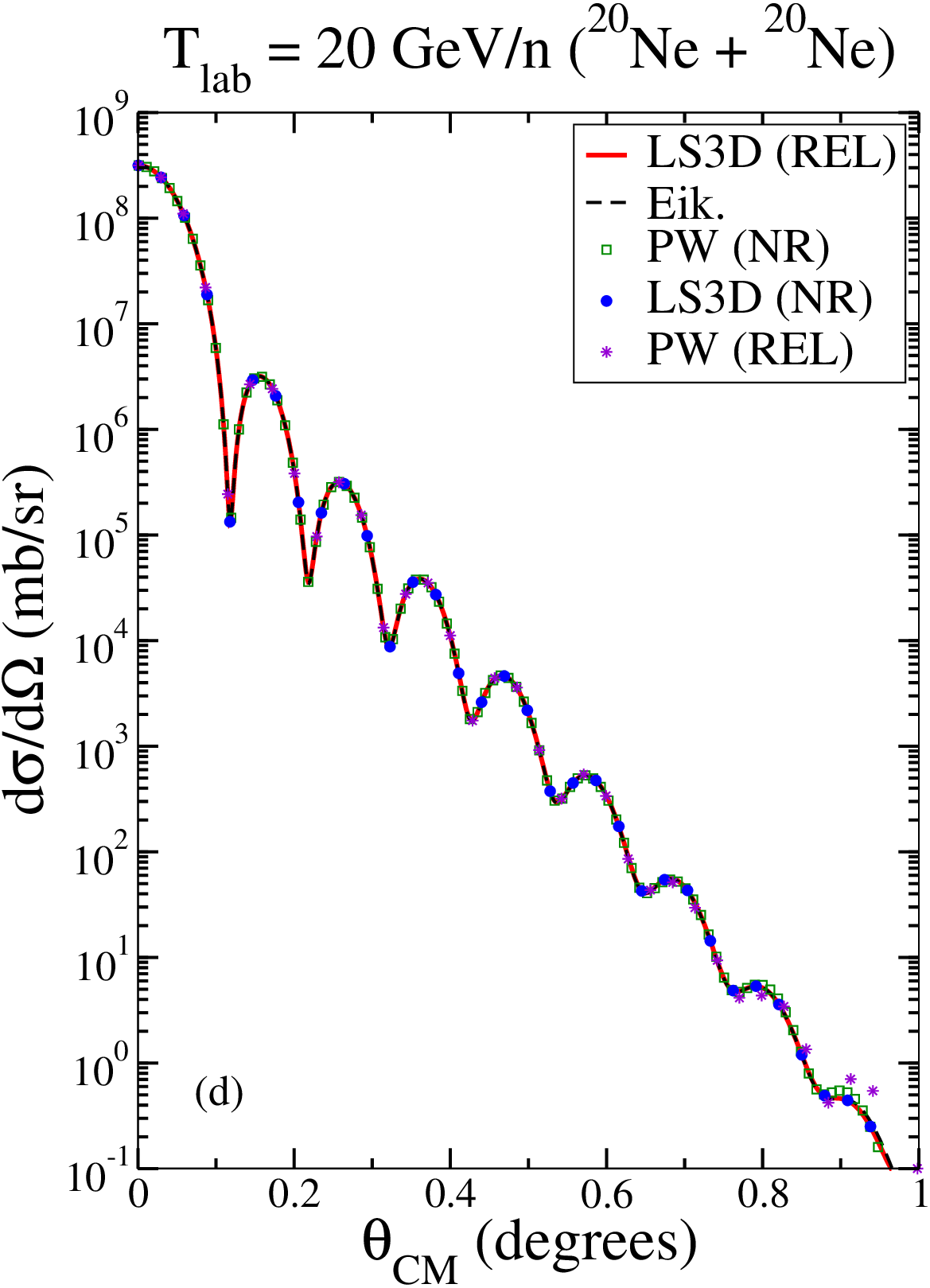}} 
\caption[Elastic differential cross sections for $^{20}$Ne + $^{20}$Ne reactions]{Elastic differential cross sections for $^{20}$Ne + $^{20}$Ne reactions for projectile lab kinetic energies of 150, 500, 1000, 20000 MeV/n.
Eik. represents eikonal, LS3D represents three-dimensional Lippmann-Schwinger, and PW represents partial wave. Non-relativistic results are denoted (NR) and relativistic results are denoted (REL).
}
\label{NeNefig}
\end{center}
\end{figure*}

\begin{figure*}
\begin{center}
\subfigure{\includegraphics[scale=0.33]{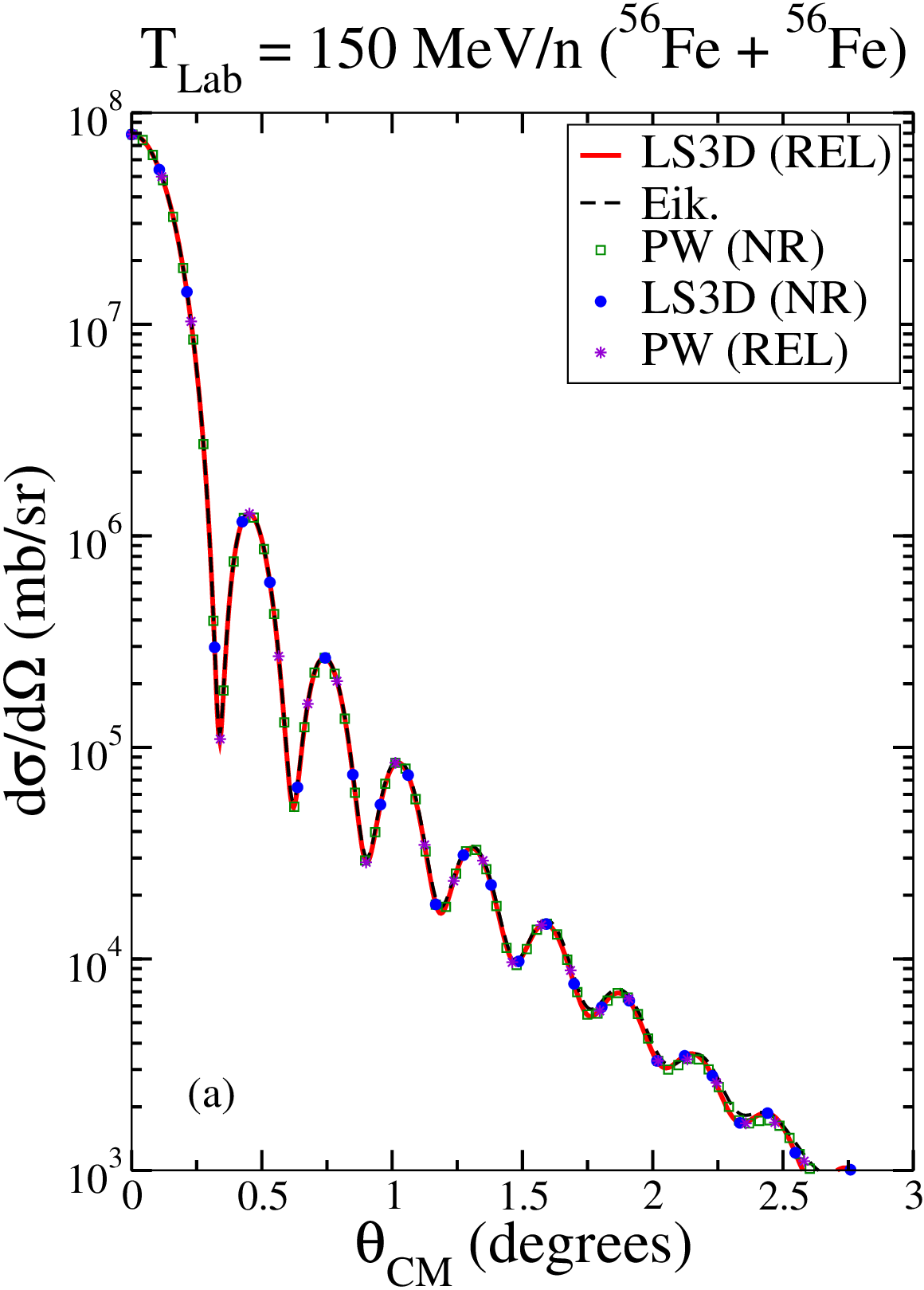}} \hspace{0.5cm}
\subfigure{\includegraphics[scale=0.33]{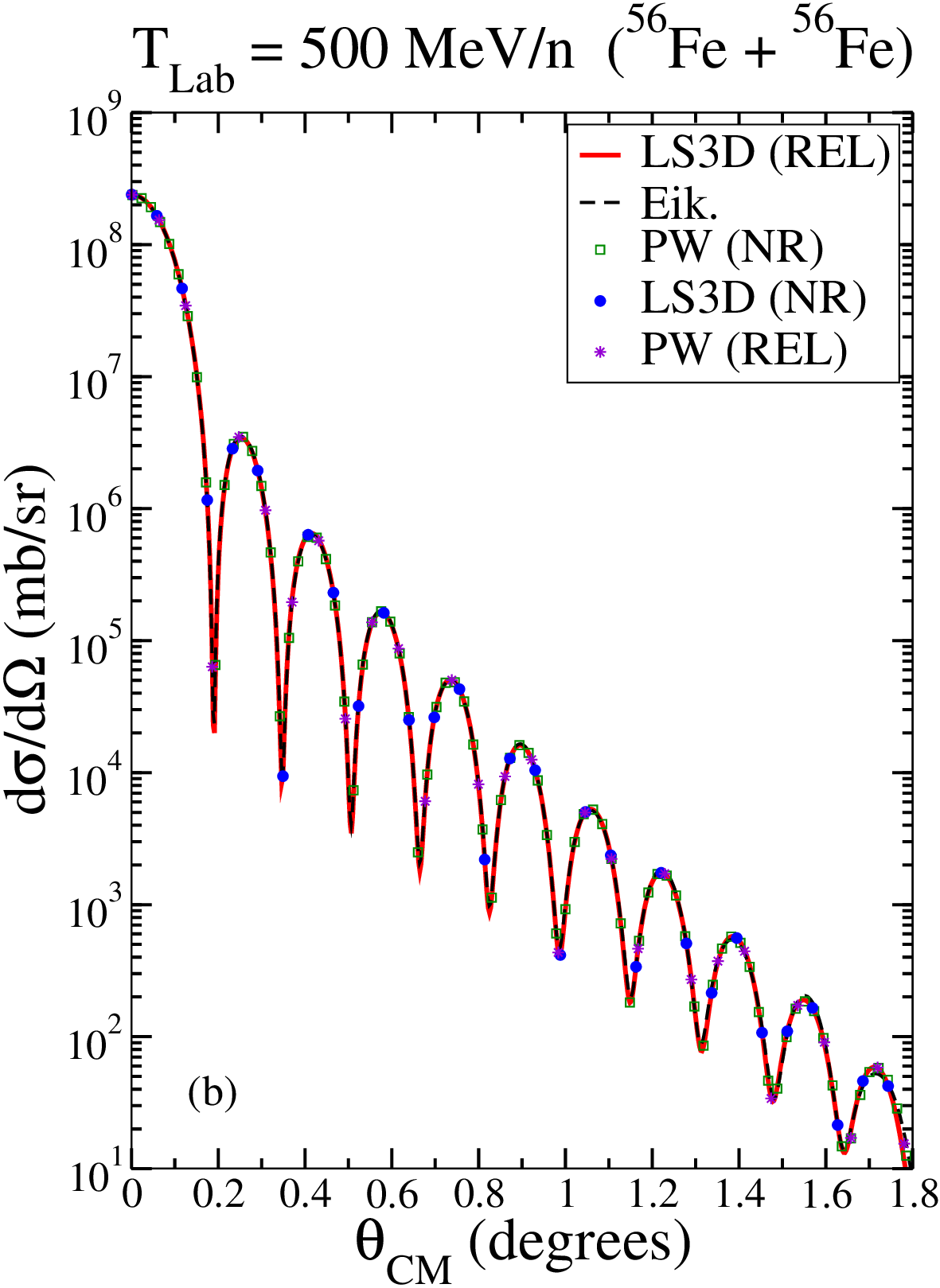}}  \\
\subfigure{\includegraphics[scale=0.33]{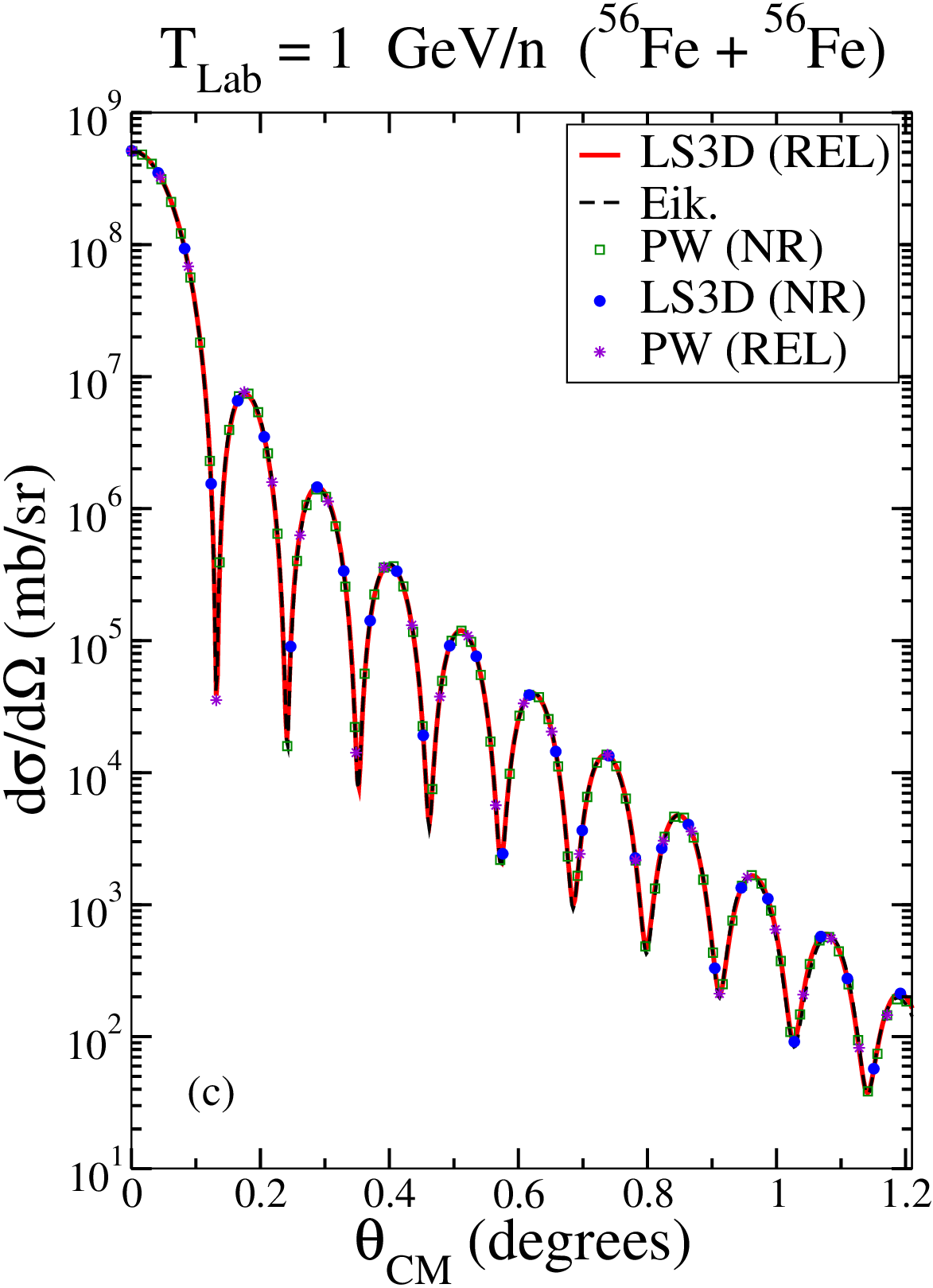}} \hspace{0.5cm}
\subfigure{\includegraphics[scale=0.33]{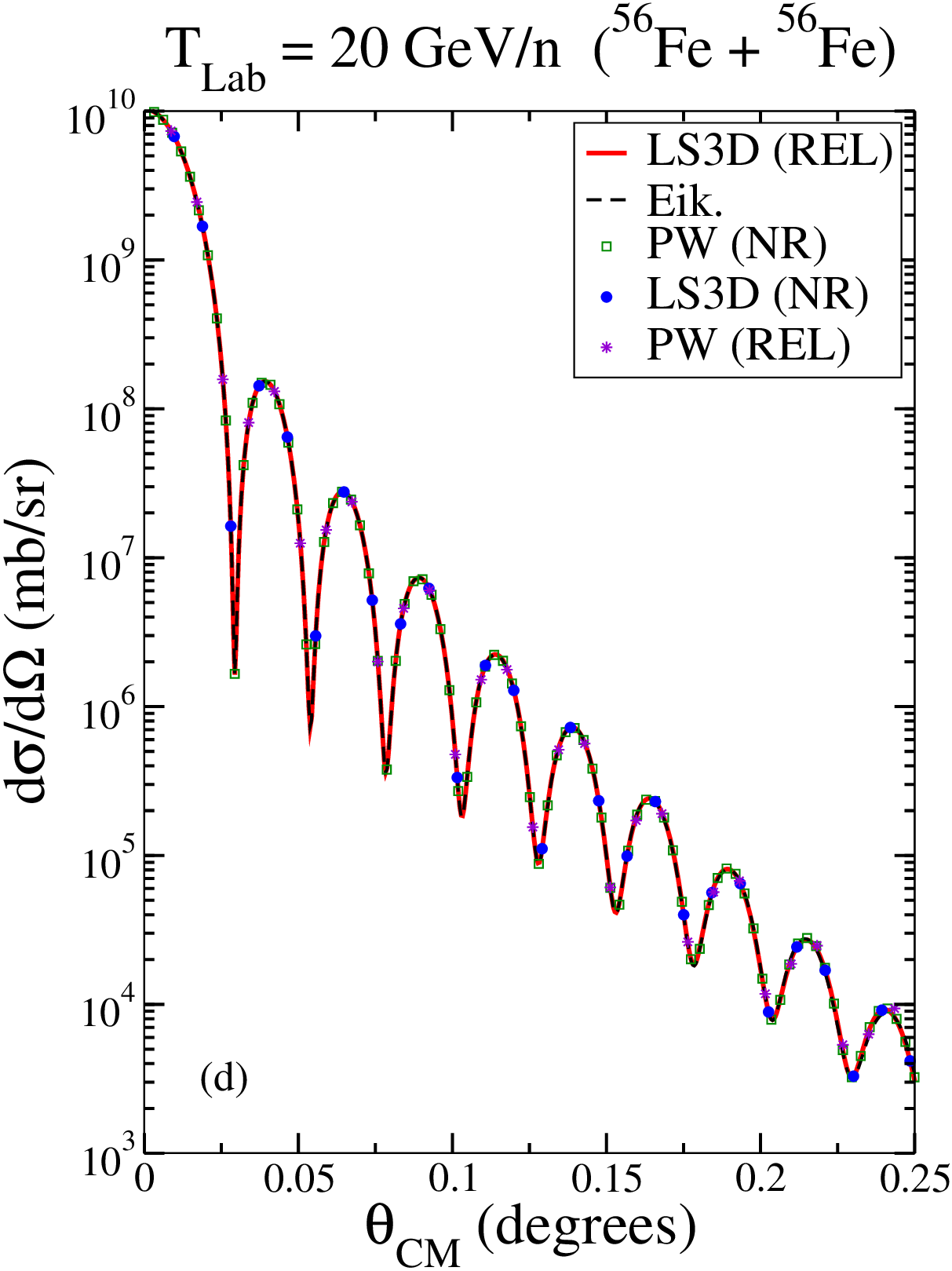}} 
\caption[Elastic differential cross sections for $^{56}$Fe + $^{56}$Fe reactions]{Elastic differential cross sections for $^{56}$Fe + $^{56}$Fe reactions for projectile lab kinetic energies of 150, 500, 1000, 20000 MeV/n.
Eik. represents eikonal, LS3D represents three-dimensional Lippmann-Schwinger, and PW represents partial wave. Non-relativistic results are denoted (NR) and relativistic results are denoted (REL).
}
\label{FeFefig}
\end{center}
\end{figure*}

Next, eikonal (Eik), partial wave (PW), and LS3D codes are used to predict the elastic differential cross sections for $^{20}$Ne + $^{20}$Ne reactions in Fig. \ref{NeNefig} and $^{56}$Fe + $^{56}$Fe reactions in Fig. \ref{FeFefig} with lab projectile kinetic energies of 150, 500, 1000, and 20000 MeV/n.  These solution methods are fully described in references \cite{Joachain, LS3D_1,LS3D_2,LS3D_3,LS3D_4,Pickle, Pickle2, Wolfe, NASATP2014}. The PW and LS3D results are generated with NR and REL kinematics, whereas the Eik code is NR. Each figure shows that there is no significant difference between the NR and REL elastic differential cross sections, regardless of the energy.

In summary, it is noted that the REL propagator has pole structure that is similar to the NR case and that the REL and NR scattering amplitudes are approximately equal near the elastic cut. The optical potential is largest near the on-shell momentum and decays rapidly thereafter. Consequently,  
the NR and REL on-shell scattering amplitudes have been shown to be nearly indistinguishable for projectile and target nuclei of equal mass. 

\section*{Acknowledgments}
The authors would like to thank Drs. John Norbury, Steve Blattnig, Ryan Norman, Jonathan Ransom, and Francis Badavi for reviewing this manuscript. 
This work was supported by the Human Research Program under the Human Exploration and Operations Mission Directorate of NASA and NASA grant number NNX13AH31A.

%\bibliographystyle{prsty}
%\bibliography{LS3Dbib}

\newpage

\clearpage

\end{document}